\documentclass{aa}
\input psfig.sty
\input epsf.sty

\def \h5-1 { $h_{50}^{-1}$}

\begin{document}
   \thesaurus{     
               11.03.1;  
               11.09.1;  
               11.07.1;  
               12.03.3;  
               12.12.1;  
               13.25.2  
} 

   \title{The NGC4839 group falling into the Coma cluster observed by 
      XMM-Newton 
\thanks{Based on observations obtained with XMM-Newton, an 
      ESA science mission with instruments and contributions directly funded by
    ESA Member States and the USA (NASA)}
\fnmsep
\thanks{EPIC was developed by the EPIC Consortium led by the Principal
Investigator, Dr. M. J. L. Turner. The consortium comprises the
following Institutes: University of Leicester, University of
Birmingham, (UK); CEA/Saclay, IAS Orsay, CESR Toulouse, (France);
IAAP Tuebingen, MPE Garching,(Germany); IFC Milan, ITESRE Bologna,
IAUP Palermo, Italy. EPIC is funded by: PPARC, CEA, CNES, DLR and ASI
}}

\titlerunning{The NGC4839 group falling into the Coma cluster observed by 
      XMM-Newton}


   \author{D.M. Neumann \inst{1}
\and
 M. Arnaud \inst{1}
\and 
R. Gastaud \inst{2}
\and
 N. Aghanim \inst{3}
\and 
 D. Lumb \inst{4}
\and
U.G. Briel \inst{5}
\and
 W. T. Vestrand \inst{6}
\and
  G.C. Stewart \inst{7}
\and
 S. Molendi \inst{8}
\and         
 J.P.D. Mittaz \inst{9}
          }

   \offprints{D.M. Neumann, email: ddon@cea.fr}

   \institute{CEA/DSM/DAPNIA Saclay, Service d'Astrophysique,
   L'Orme des Merisiers B\^at. 709., 91191 Gif-sur-Yvette, France
\and
CEA/DSM/DAPNIA Saclay, SEI, 91191 Gif-sur-Yvette, France
\and
IAS-CNRS, Universi\'e Paris Sud, B\^atiment 121, 91405 Orsay Cedex, France
\and 
   ESTEC, European Space \& Technology Centre, Keplerlaan 1, Postbus 1,
2200 AG Noordwijk, the Netherlands
\and
   Max-Planck Institut f\"ur extraterrestrische Physik, 
 Giessenbachstr.,
85740 Garching,
Germany
\and
  NIS-2, MS D436,
  Los Alamos National Laboratory,
  Los Alamos, NM, 87454  USA
\and
Dept. Of Physics and Astronomy, Leicester University,
Leicester LE1 7RH, UK
\and
IFC/CNR, Via Bassini 15, I-20133, Milano, Italy
\and
University College London,
              Mullard Space Science Laboratory,
              Holmbury St. Mary,
              Nr. Dorking,
              Surrey,
              RH5 6NT,
              U.K.
                 }

  \date{Received September 29, 2000; accepted October 18, 2000}

   \maketitle

   \begin{abstract}
We present here the first analysis of the XMM Newton EPIC-MOS data of the 
galaxy group around \object{NGC4839}, which lies at a projected distance to 
the \object{Coma cluster} center of 1.6\h5-1 ~Mpc. In
our analysis, which includes imaging, spectro-imaging and spectroscopy
we find compelling evidence for the sub group being on its first infall onto 
the Coma cluster. The complex temperature structure around NGC 4839 is 
consistent with simulations of galaxies falling into a cluster environment. We 
see indications of a bow shock and of ram 
pressure stripping around NGC4839. Furthermore our data reveal a 
displacement between NGC4839 and the center of the hot gas in the group of 
about 300 \h5-1 ~kpc. With a simple approximation  we can explain this 
displacement by the pressure force originating from the infall, which acts 
much stronger on the group gas than on the galaxies.
   \end{abstract}

%

\section{Introduction}

There is now compelling evidence from studies in different wavelength bands
that the Coma cluster has substructure. The search for substructure in the
Coma cluster starts more than a decade ago: Fitchett \& Webster
(\cite{fitchett}) find a double structure in the cluster center, and Mellier 
et al. (\cite{mellier}) find a 
second peak in the large scale galaxy distribution around NGC4839, which 
was further confirmed in a study by Merritt \& Tremblay (\cite{merritt}). 
While these
studies were based on optical data, Briel et al. (\cite{briel1})
using X-ray data coming from the all-sky-survey of the ROSAT satellite, found 
substructures coinciding with NGC4911, and again, NGC4839. Another
study by White et al. (\cite{white}) based on a deeper ROSAT 
observation 
revealed even more structure. In this paper we concentrate on the XMM-Newton
(Jansen et al. \cite{jansen}) 
observation of the sub group around NGC4839, which has a projected distance 
from the Coma cluster of 1.6 \h5-1 ~Mpc. In section 2 we review briefly the
observation and data treatment of the EPIC data (see Turner et al. 
\cite{turner}). 
This is followed by Sections dealing 
with  imaging
and spectro-imaging. In section 5 we perform a spectroscopic analysis in 
different regions which is followed by discussion and conclusion.
This paper is one of three papers which present XMM-Newton data on the
Coma cluster (see also Arnaud et al. \cite{arnaudm1}; Briel et al. 
\cite{briel2}).
 
Throughout this paper, we assume $H_0=50$~ km s$^{-1}$ Mpc$^{-1}$ \h5-1 and
$q_0 = 1/2$. At a redshift of $z=0.0232$,  $1^\prime$ corresponds to 38.9~kpc. 


\section{Observations}

The subgroup associated with NGC4839 was observed for 36~ksec with XMM-Newton 
in
full frame mode with the medium filter. Due to remaining problems with the CTI
(Charge Transfer Inefficiency) 
calibration in the extended full frame mode of the PN camera
we concentrate in this paper on the data from the EPIC-MOS1 and EPIC-MOS2 
cameras.
The background intensity
of the instruments varies considerably with time over several
orders of magnitude. In order to optimize the signal-to-noise of the 
observation we reject time intervals with high background count rates. For this
selection we binned the counts  in the energy band
10-12~keV in time intervals of 100~sec. In this energy band we hardly detect
any astrophysical sources, due to the very small effective area of the mirrors.
 We reject time bins with more than 15 counts per 100~sec.  
After this selection we obtain exposure times with 
30.6~ksec for EPIC-MOS1 and 30.9~ksec for EPIC-MOS2. 

\section{Imaging}

\begin{figure}
\psfig{figure=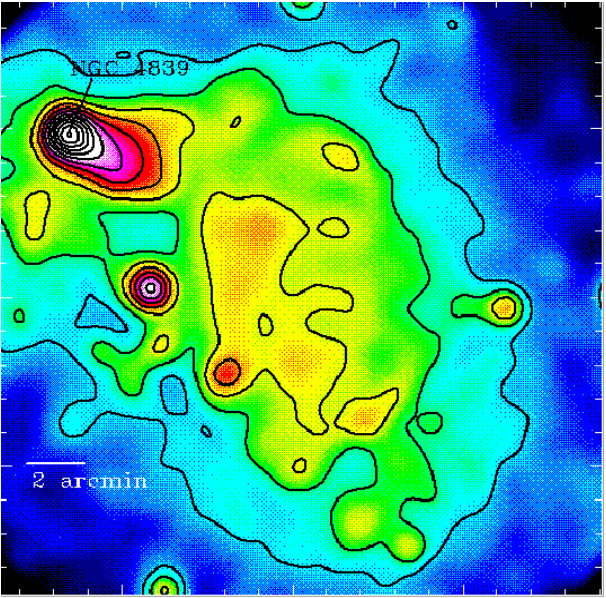,width=8cm}
\caption{The 0.5-2.0~keV image of the EPIC-MOS1 and EPIC-MOS2 
camera. To correct for 
vignetting we corrected each individual photon with a weighting function
(see Arnaud et al. \cite{arnaudm2}).
The image is gauss filtered with $\sigma=30^{\prime\prime}$. The spacing of
the contours is linear with a step size of $1.17\times 10^{-3}$
cts/sec/arcmin$^2$. The lowest
contour is at $4.68\times 10^{-3}$  cts/sec/arcmin$^2$. }
\label{fig:im05_2}
\end{figure}

\begin{figure}
\psfig{figure=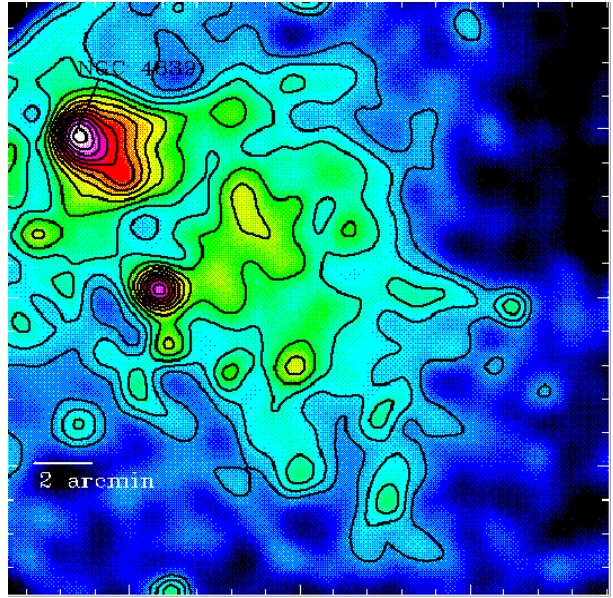,width=8cm}
\caption{The 2.0-5.0~keV image of the EPIC-MOS1 and EPIC-MOS2 camera. To 
correct for 
vignetting we corrected each individual photon with a weighting function
(see Arnaud et al. \cite{arnaudm2}).
The image is gauss filtered with $\sigma=30^{\prime\prime}$. The spacing of
the contours is linear with a step size of $2.34\times 10^{-4}$ 
cts/sec/arcmin$^2$. The lowest
contour is at $1.87 \times 10^{-3}$ cts/sec/arcmin$^2$. }
\label{fig:im2_5}
\end{figure}

Fig.\ref{fig:im05_2} and Fig.\ref{fig:im2_5} display the images obtained in 
the band 0.5-2.0~keV and 2.0-5.0~keV, respectively.
The band 5.0-10.0~keV is dominated by the particle background. 
 As already observed with ROSAT (Briel et al. \cite{briel1}; 
White et al. \cite{white}; Dow \& White \cite{dow}) we see a tail-like 
structure around NGC4839 pointing away from the Coma cluster. The orientation 
of NGC4839 in
the optical 
coincides with the orientation of the tail. The X-ray emission of NGC4839 
seems somewhat 
disconnected to the diffuse X-ray emission coming from the group in the center 
of the pointing.

\section{Spectro-imaging}

\begin{figure}
\psfig{figure=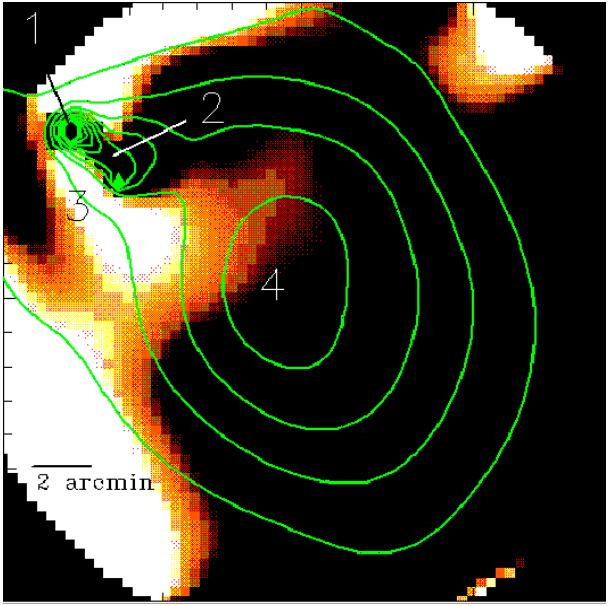,width=8cm}
\caption{The background subtracted hardness ratio image in the bands 2.-5.0~keV
and 0.5-2.0~keV. The image shows the data only up to a radius of 12~arcmin.
Regions outside this area do not show sufficient statistics to give meaningful
results. The contours are the results from a wavelet filtering of an image
in the band 0.5-2.0~keV.}
\label{fig:hr}
\end{figure}

In order to look for spectral variations we calculated a hardness ratio image
of the data (see Fig.\ref{fig:hr}). To account for the background, especially 
the particle background,
we used another Coma pointing with the central coordinates of RA=13h01min37sec
Dec=27d19m52s (J2000), which
is situated South-East of the Coma cluster at roughly the same offset angle
than the sub-group.
(46 arcmin in comparison to 51 arcmin for the subgroup pointing).
We orientate this background pointing in such a way that it accounts for the
emission of Coma itself plus stray light in the pointing for NGC4839. However,
this rotation has no measurable effect on the hardness ratio image, the 
features with or without this rotation rest identical. This indicates that
the background we measure is dominated by particles.

As we are not interested here in
the emission of different point sources but only in spectral differences of
the extended X-ray emission we use a wavelet algorithm (the MVM algorithm
Ru\'e \& Bijaoui \cite{rue}) to filter out point-like
structures for the hardness ratio images. Our hardness ratio image is 
calculated as hr = (image(2-5~keV)-image(0.5-2.0~keV))/
(image(2-5~keV)+image(0.5-2.0~keV)).

We do not take into account any vignetting effects in the hardness ratio image
since we know from calibrations that the vignetting functions are parallel 
to each other for energies below 5~keV.

The resulting image is shown in Fig.\ref{fig:hr}. 
We can see that there are indeed spectral variations across the field-of-view.
Around NGC4839 we see that the tail-like structure has a different colour 
(is softer) than the regions North or South  surrounding it 
(the tail has a mean hardness ratio of $\langle$hr$\rangle=-0.6$). 
The region, which shows a harder spectrum in the South 
($\langle$hr$\rangle=-0.4$) seems also to have a
tail-like structure, which is parallel to the tail seen in the original image. 
The core of NGC4839 shows $\langle$hr$\rangle=-(0.5-0.6)$

The appearance of a region with hard spectrum in the South-East or North, at 
the border of the hardness ratio image is not statistically significant. We 
cannot state with certainty that these regions are hotter due to an interaction
of the group with the intracluster medium (herafter ICM) of the Coma cluster. 
These ``hot'' regions
are not unlikely to be an artifact which is created by low photon statistics 
together with remaining stray light. 

\section{Spectroscopy}

\begin{figure}
\psfig{figure=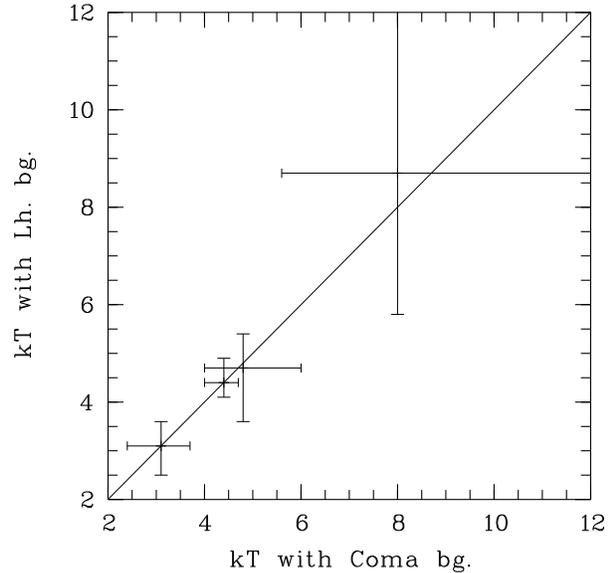,width=8cm}
\caption{The temperature estimates of the four regions with different 
background observations. The horizontal axis: Coma background, the vertical 
axis: Lockman hole background. Error bars are 90\% confidence level.
}
\label{fig:comp}
\end{figure}

\begin{figure*}
\begin{tabular}{cc}
\psfig{figure=XMM09_f5.ps,width=8.5cm}&
\psfig{figure=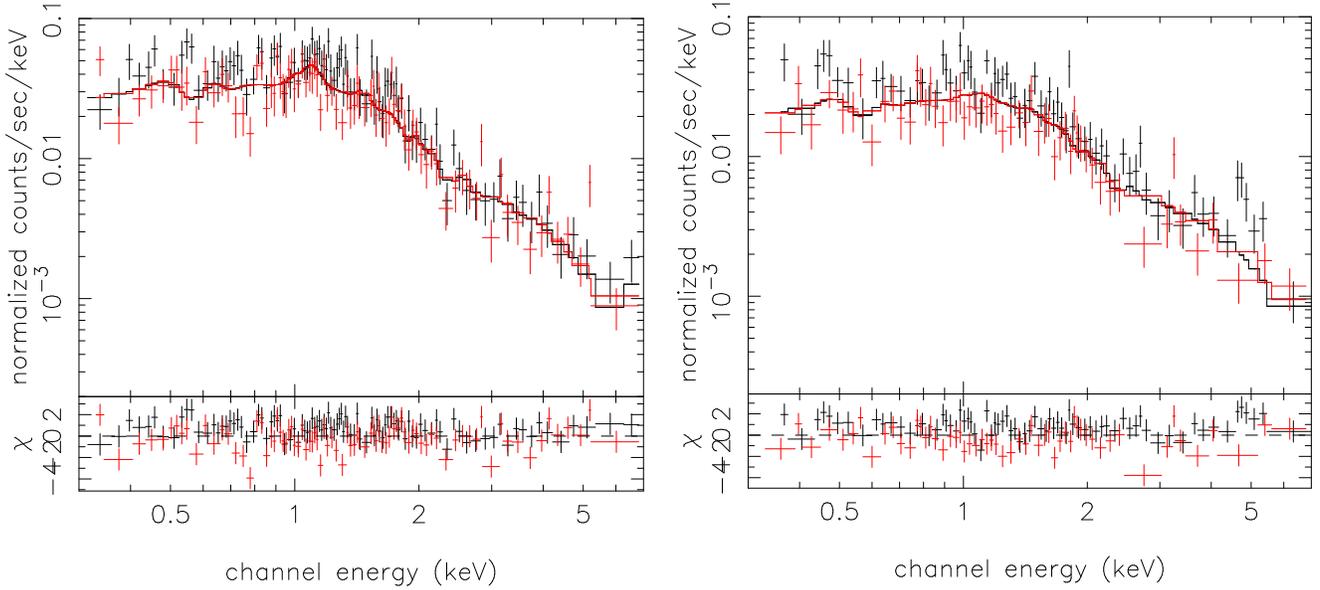,width=8.5cm}
\end{tabular}
\caption{The spectra of region 2 (left) and region 3 (right). The black
crosses show EPIC-MOS1 data, the red crosses show EPIC-MOS2 data. }
\label{fig:spec}
\end{figure*}

For our spectral analysis we selected different regions in the pointing, which
show spectral differences in the hardness ratio image. In our spectral 
analysis we restrict ourselves to the energy range between 0.3--10.0~keV.
The summary of the spectral fitting is given in Tab.\ref{tab:spec}. The
position and extraction regions are displayed in Tab.\ref{tab:reg}. The
spectra for some of the regions are shown in Fig.\ref{fig:spec}.
For the spectral fitting we used XSPEC (Arnaud \cite{arnaudk})
and the MEKAL code (Mewe et al. \cite{mewe1}; Mewe et al. \cite{mewe2}; 
Kaastra \cite{kaastra};
Liedahl et al. \cite{liedahl}). The galactic absorption was 
fixed to the value of the 21cm line ($nH=9\times 10^{19}$cm$^{-2}$  -- Dickey
\& Lockman \cite{dickey}).
For the background spectrum determination we used the observations performed 
on the Coma background pointing (see above). 
In order to take into account instrumental variations across the field-of-view
we selected for the background the same regions on the detector than for the 
source spectra. As the particle contribution is somewhat detector dependent,
it is better to take the same regions on the detector rather
than the regions corresponding to the Coma emission (the orientated regions,
like for the hardness ratio image), since
our background is highly dominated by particles and not by cluster 
emission/stray light.
In order to estimate the influence of background variations
we also used the observations of the Lockman hole as background. For EPIC-MOS1 
we used the Lockman hole observations with the thin filter, which is not 
entirely consistent, since all Coma pointings were observed with the
medium filter.
However, this will give us an estimate on the intrinsic variations of the
background. The comparison of the best fit temperature estimates are shown in
Fig.\ref{fig:comp}. As one can see, the temperatures are very similar.

To account for vignetting we used the approach described in Arnaud et al.
(\cite{arnaudm2}).

\begin{table}
\begin{tabular}{cccccc}
\hline
Region \hspace*{-0.4cm}
& description & $kT$ \hspace*{-0.2cm}
& metal & $\chi^2$red. \hspace*{-0.4cm}& $\Gamma$ \\
& & & Abund. & & \\
\hline
\hline
1 & core of & $3.1_{2.4}^{3.7}$ &  $0.56_{0.28}^{0.96}$ & 0.82 &
$1.78^{1.89}_{1.69}$ \\
& NGC4839 &  & & & \\
2 & tail of  & $4.8_{4.0}^{6.0}$ & $0.70_{0.31}^{1.6}$ & 0.99 & \\
 &  NGC 4839 & & & & \\
3 & region south  & $8.0_{5.6}^{13.}$ & $0.01_{0.0}^{2.8}$ & 0.69 &
$1.51_{1.43}^{1.62}$ \\
 & of tail & & & & \\
4 & main body of  & $4.4_{4.0}^{4.7}$ & $0.36_{0.26}^{0.50}$ & 0.87  &\\
 &  sub group & & & \\
\hline
\end{tabular}
\caption{Summary of the spectral fits in different regions for the combined 
spectra of EPIC-MOS1 and EPIC-MOS2. The errors are
90\% confidence level. In order to account for background variations we allowed
a 10\% error in the normalization of the background spectrum for MOS1 and MOS2.
For the fitting of a thermal plasma we used the MEKAL code in XSPEC. 
}
\label{tab:spec}
\end{table}

\begin{table}
\begin{tabular}{cccc}
\hline
region & center R. A.& center Dec. & radius \\
       & (h:min:sec) & (deg:arcmin:arcsec) & in arcsec\\
\hline
\hline
1& 12:57:24 & 27:29:54 & 40 \\
2& 12:57:17 & 27:29:37 & 53 \\
 & 12:57:08 & 27:29:06 & 65 \\
3 & 12:57:23 & 27:27:50 & 68 \\
  & 12:57:15 & 27:26:49 & 63\\
4 & 12:56:53 & 27:24:27 & 205\\
\hline
\end{tabular}
\caption{Definition of the circles for the spectra. The coordinates are in 
J2000. For region 2 and 3 we used two circles for extracting the spectra.}
\label{tab:reg}
\end{table}

\subsection{Region 1: core of NGC4839}

The origin of this X-ray emission is somewhat unclear. It might be the AGN
of NGC4839, which is known as a radio source, or the emission might be thermal.
Fitting
a power law with the results as shown in Tab.\ref{tab:spec} gives a reduced
$\chi^2$ in the order of 1.3. Fitting a thermal plasma improves the fit with
a reduced $\chi^2$ of about 0.8-0.9.

\subsection{Region 2: the tail of NGC4839}

This region, for which we only adopted a thermal emission model shows high
temperature and metallicity. The temperature is too high for a galaxy, which
ranges more about 1~keV and not around 4.5~keV, as observed. This might be
explained by ram pressure stripping: the hot gas is blown out as the medium 
encounters the ICM of the Coma cluster. Due to the high velocity of the infall
of NGC 4839 into the Coma cluster, shocks could occur, which heat the 
intergalactic medium. The fact that we see the tail pointing away from the 
center of the Coma cluster is a strong indication for this scenario (see
also discussion).

\subsection{Region 3: the region South of NGC4839}

As can be seen in Fig.\ref{fig:hr}, the region South of NGC4839 seems to have 
a different spectrum with respect to center or tail of NGC4839.
We fit a thermal plasma and a power law to the spectrum of this region. The
results, again are shown in Tab.\ref{tab:spec}.
Both models for the emission are acceptable. The intergalactic medium around
NGC4839 could be heated by shocks caused from the infall of NGC4839 onto the 
Coma cluster (see also the discussion section).

A power law emission could be linked to the particles
responsible for the radio emission observed in NGC4839. 
NGC 4839 is a radio source showing two opposite jets and a complex
radio tail.  This radio tail is extended 
in S-W  direction and coincides nicely
 with region number 3 (see Venturi et al. \cite{venturi}). If 
the X-ray emission in this 
region is linked to the radio tail it could come from Inverse Compton 
scattering of cosmic microwave background photons as they interact with the 
relativistic electrons originating from the radio jet. However, in this case it
is not clear whether XMM has intrinsically enough sensitivity to resolve this 
IC emission.

Comparing the surface brightness of region 2 and region 3 we find a difference 
in intensity of a factor of about 3.5 in the energy band $0.5-2.0$~keV. If we 
assume that the emission is thermal and thus the radiated energy $E\propto n^2$
and furthermore that region 2 and 3 are in pressure equilibrium we expect a 
difference in surface brightness of the order of 2.5, since region 2 has about 
$kT$=5~keV and region 3 has most likely a temperature of about $kT$=8~keV. The 
remaining difference between the surface brightness intensities might, in this 
case be due to differences in metallicities or extend in the line-of-sight. 
Region 2 has a high metallicity and is in a temperature range, where
line emission is more dominant than at 8~keV. 

Taking this all together (see also discussion) we think that it is more likely 
that the emission of region 3 is due to thermal bremsstrahlung rather than due
to IC scattering.

\subsection{Region 4: main body of sub group}

This is the center of the extended X-ray emission of the subgroup. The 
fitted temperature of about $kT$=4.4~keV is consistent 
with ASCA results by Watanabe et al.(\cite{watanabe}). The metallicity of this 
group seems 
to be typical with a best fit abundance of 0.3. The abundance seems to be 
significantly smaller than the one observed in the tail of NGC4839 (region 2).

\section{Discussion}

\subsection{The structure around NGC 4839}
There have been already several studies dealing with the physics of
galaxies falling onto clusters, like Gaetz et al. (\cite{gaetz}), 
Balsara et al. (\cite{balsara}) or 
Stevens et al.  (\cite{stevens}) (hereafter SAP99) and references 
therein.
SAP99 found different dynamical features for the infall depending on
cluster gas temperature, galaxy mass replenishment rate and infall velocity.
SAP99 performed simulations based on a 2D hydrodynamic code, which is similar
to the one used in Balsara et al. (\cite{balsara}). The ICM follows
a beta-model distribution with $\beta=2/3$ and a mass distribution of the
 which extends up to 32.2~kpc. Similar parameters for the galaxy
distribution have been used by Gaetz et al. (\cite{gaetz}).     
The hardness ratio image Fig.\ref{fig:hr} around NGC 4839 shows morphological
similarities to their model 2c. The physical properties of this model 2c are 
an intermediate cluster at 4~keV and an infall velocity of 1380~km/sec. These
numbers are slightly different to what is observed in the case of NGC4839.
Coma has a gas temperature of about 8~keV and an infall velocity of 1700~km/sec
(Colless \& Dunn \cite{colless}, hereafter CD96). Despite the  differences in 
actual numbers  we 
interpret the striking resemblances as a strong indication for NGC4839 being
on its first infall onto the Coma cluster (see for example CD96), and not, as 
suggested by Burns et al. (\cite{burns}) going out of the cluster
after having already passed through the Coma center.

\subsection{Displacement of NGC4839 and the extended X-ray emission}

Another feature which is striking in our observations is the displacement
of the extended X-ray emission belonging to the sub group of NGC 4839 and NGC
4839 itself of about 8~arcmin (see for example Fig.\ref{fig:hr}). This 
corresponds to roughly 300$h_{\mathrm 50}^{-1}$~kpc. 
From the optical study by CD96
NGC 4839 seems to be right in the centroid of the galaxy distribution 
belonging to the 
sub group. It seems therefore likely that the galaxies belonging to the
sub group are on average closer to the Coma center than the hot gas belonging 
to the sub group. Such a displacement between galaxies and gas is expected when
the sub group encounters the gas of the main cluster. The 
ICM of the main cluster slows down the infall. As the galaxies in the
sub group show larger densities and therefore a smaller surface-to-mass ratio
than the gas, the galaxies are less affected by the ram pressure force than the
intra group gas. In the following we make a first order calculation
to
see whether this effect is large enough to explain the displacement of about
300$h_{50}^{-1}$~kpc between NGC4839 and the center of the gas. For a rough
calculation we estimate that the acceleration due to gravitational forces is
much stronger than the counteracting ram pressure force. We furthermore assume 
a total mass of the Coma cluster of $2\times 10^{15}h_{50}^{-1}$M$_\odot$
(Hughes \cite{hughes}; Briel et al. \cite{briel1}; CD96),
and that the Coma cluster mass is a factor of ten higher than the mass of the 
infalling group (CD96). In a very rough approximation we suppose that the Coma
 mass 
distribution is to first order point-like. This gives an infall velocity 
$v=\sqrt{2GM/r}$ of NGC4839 of 2400~km/sec, supposing a distance to the Coma 
cluster to $r=1.6$~Mpc (this is the projected distance see also 
Dow \& White \cite{dow}, or
CD96 -- the most likely infall direction is close to the plane-of-sky). The 
approximated infall velocity is not too 
different to the calculated value of about $1700^{+350}_{-500}$~km/sec by 
CD96. 
For the ram pressure we use $P_{\mathrm ram}=\rho v^2$  and for the density 
distribution of
the ICM a beta-model ($\rho = \rho_0(1+r^2/r_{\mathrm c}^2)^{-3\beta/2}$). For 
simplicity
we assume $\beta=2/3$ and $r_{\mathrm c}=400h_{50}^{-1}$~kpc, which is in good
agreement with the ROSAT measurements by Briel et al. (\cite{briel1}). 
For $n_{\mathrm e0}$ we use their calculated value of $3\times 
10^{-3}h_{50}^{-1/2}$ 
cm$^{-3}$ and transform it into $\rho_0$.
For the surface of the gas encountering the ICM of Coma we suppose a circular
surface with a radius of $300h_{50}^{-1}$~kpc (we take the radius of the 
outermost contour on Fig.\ref{fig:hr}) and a gas mass of about 
$10^{13}$M$_\odot$.
The force exerted by the ram pressure is $P_{\mathrm ram}S=\rho v^2S=ma$. $S$ 
is 
the 
surface of the sub group gas, $m$ its mass and $a$ the deceleration.
For the slowing down we calculate $v_{\mathrm dec} = \int a dt$ $=\int 
(PS/m dt/ds) ds$
$ = S/m \int(\rho v_{\mathrm infall})ds$ . We suppose $r^2>>r_{\mathrm c}^2$, 
which is 
certainly true at 
distances greater 1.6~Mpc. Filling in all parameters we get 
$v_{\mathrm dec}= 540$~km/sec, which means that the actual infall velocity is 
diminished by 540~km/sec for the hot gas of the sub group. Extrapolating a 
difference in velocity between NGC4839 and
the gas of 540~km/sec (we assume that the ram pressure force is negligible for
the galaxy) to the time which the group needed to go from the virial radius 
(about 3~Mpc) to the observed distance of 1.6~Mpc we see that we obtain a 
difference in location of about 400~kpc assuming a constant infall velocity of
2000~km/sec for NGC4839. This value which was determined with many 
approximations is of the same order of magnitude as the observed displacement
between gas and galaxy. The differences in location
between NGC 4839 and the group gas is thus likely to be due to the ram 
pressure 
exerced by the pressure of Coma's ICM. In addition, 
more detailed modeling of 
this infall is certainly very interesting, since it is very likely that 
the displacement can be used to determine the density
or temperature of the ICM at large radii.

\subsection{The mass ratio NGC4839 group - Coma cluster}

We obtain a temperature of the sub group of about 4.5~keV. Comparing this to
the central temperature of the Coma cluster of about 8--9~keV and applying
the M-T relation of $M \propto T^{3/2}$ we obtain a mass ratio of 
$M_{\mathrm NGC4839}/M_{\mathrm Coma} \sim 0.4$. This ratio is very high in 
comparison 
to the 
result of CD96, who find that the sub group has 5-10\% the mass of the Coma
cluster. This discrepancy can be either explained by superposition of the
gas in the group with the hotter ICM of the Coma cluster or  
the fact that the ICM of the 
NGC4839 group is heated up above its initial virial temperature. A plausible
scenario is that the subgroup is in a process of internal merging. -- The
somewhat disrupted morphology of the extended emission seen
in Fig.\ref{fig:im05_2}
and Fig.\ref{fig:im2_5} is an indication for that.
Such processes of internal merging
were  already observed in hydrodynamic simulations (see for 
example Schindler \& M\"uller \cite{schindler}). If the ICM of the sub group 
were heated 
up by a factor of two, which is reasonable, and its initial virial temperature 
was 2~keV, we would obtain a mass ratio of $M_{\mathrm NGC4839}/M_{\mathrm 
Coma} \sim
(2/8.5)^{3/2} = 0.11$, which is in good agreement with the values obtained by
CD96.

\section{Conclusion}

We performed a first analysis of the XMM-Newton data of the NGC4839 group,
which has a projected distance of the Coma cluster of 1.6$h_{50}^{-1}$~Mpc.
In our analysis we resolve for the first time a complex temperature structure 
around NGC4839 which can be explained by a first infall of the galaxy onto the 
Coma cluster. 

Furthermore, we measure a displacement between NGC4839, the dominant galaxy of
the group and the extended X-ray emission coming from its ICM. In a simple
approximation we can explain this displacement by a ram
pressure force of Coma's 
ICM acting on the hot gas of the sub group as it falls onto the Coma center.

\begin{acknowledgements}
We would like to thank J. Ballet for support concerning the SAS software and
J.-L. Sauvageot for providing the gain correction. We want to thank  
L. Feretti and A. Bykov for very useful discussions.
\end{acknowledgements}

\end{document}